\documentclass[twocolumn, twocolappendix]{aastex63}

\usepackage[utf8]{inputenc}
\usepackage{graphicx}
\usepackage{threeparttable}
\usepackage[caption=false]{subfig}
\usepackage{amsmath}
\usepackage{bm}
\usepackage{placeins}
\usepackage{appendix}
\usepackage{color}

\newcommand{\pwn}{G21.5$-$0.9}

\begin{document}
\title{Magnetic Field Structure and Faraday Rotation of the Plerionic Supernova Remnant \pwn\ }

\author[0000-0003-3601-5127]{Paul C. W. Lai}
\affiliation{Department of Physics, The University of Hong Kong, 
  Pokfulam Road, Hong Kong}

\author[0000-0002-5847-2612]{C.-Y. Ng}
\affiliation{Department of Physics, The University of Hong Kong, 
  Pokfulam Road, Hong Kong}

\author[0000-0002-8848-1392]{Niccolo' Bucciantini}
\affiliation{INAF – Osservatorio Astrofisico di Arcetri, Largo E. Fermi 5, I-50125 Firenze, Italy}
\affiliation{INFN – Sezione di Firenze, Via G. Sansone 1, I-50019 Sesto F. no (Firenze), Italy}
\affiliation{Dipartimento di Fisica e Astronomia, Universit\`a degli Studi di Firenze, Via G. Sansone 1, I-50019 Sesto F. no (Firenze), Italy}

\correspondingauthor{Paul C. W. Lai}
\email{paullai@connect.hku.hk}


\begin{abstract} 

We present a polarimetric study of the pulsar wind nebula (PWN) in supernova remnant \pwn, using archival Very Large Array data taken at 5 and 7.4\,GHz. The rotation measure (RM) map of the PWN shows a symmetric pattern that aligns with the presumed pulsar spin axis direction, implying a significant contribution to the RM from the nebula. We suggest that the spatial variation of the internal RM is mostly caused by the nonuniform distribution of electrons originating from the supernova ejecta. Our high-resolution radio polarization map reveals a global radial $B$-field. We show that a simple model with an overall radial field and turbulence on a small scale can reproduce many observed features of the PWN, including the polarization pattern and polarized fraction. 
The modeling results also reject a strong large-scale toroidal $B$-field, suggesting that the toroidal field observed in the inner PWN cannot propagate to the entire nebula. 
Lastly, our model predicts that the internal Faraday rotation would break the linear relation between the polarization angle and the square of the wavelength, and cause severe depolarization at low frequencies.
\end{abstract} 



\section{Introduction} 
\label{sec:inrtoduction}

A pulsar carries a strong magnetic field and rotates very rapidly. It acts as a rotating magnetic dipole,  generating a strong electric potential that accelerates electrons and positrons to relativistic speeds. The relativistic particles launched from the pulsar, together with the electromagnetic Poyting flux,  are known as pulsar wind. Upon crushing into its surroundings, e.g., supernova (SN) ejecta, it creates a termination shock (TS). The shock accelerates the particles and randomizes their motion, resulting in synchrotron radiation. Such a synchrotron bubble surrounding the pulsar is called a pulsar wind nebula (PWN). 
Shock acceleration and magnetic reconnection play an important role in accelerating the pulsar wind particles \citep{amato20}. It is found that many young pulsars have X-ray tori/rings around them \citep[see][]{ng04, ng08}.
These correspond to the pulsar wind TS location, and the shape is caused by latitude-dependent pulsar wind \citep{bogovalov+02}.


Force-free pulsar magnetosphere solutions suggest that the magnetic field is toroidal close to the pulsar light cylinder \citep{conto+99}, and this is believed to hold true before the pulsar wind reaches the TS. Theoretical work has not yet been able to describe the postshock magnetic structure in the full body of the nebula. Observationally, young PWNe show a large diversity in the global magnetic field configuration. For example, 
Vela has a toroidal field \citep{dlmd03}
and
3C58 has a $B$-field parallel to its pulsar spin axis \citep{reich02}.
The magnetic field structures of G18.95$-$1.1 and G328.4+0.2 look similar to 3C58 \citep{reich02, jm04}, but the directions of their pulsar spin axes  have not yet been identified. 
\pwn\  \citep{bs81, reich02} and DA 495 \citep{kothes+08} have large-scale radial magnetic fields, while the Crab Nebula has a radial field only near its boundary \citep{crab_mag}. 
Both \pwn\  and the Crab show toroidal fields close to the TS, as theories predicted \citep{zajczyk+12, Moran_Shearer+13b}.
Some other PWNe simply have disordered $B$-field, as in G292.0+1.8 \citep{gaensler+wallace03}. 


\pwn\ is a composite supernova remnant (SNR) consisting of a PWN core surrounded by an SNR shell. Both the PWN and the shell are highly spherical, with angular diameters of 80\arcsec\ and 5\arcmin, respectively \citep{matheson+10}.
At the center of the SNR there is a radio pulsar, J1833$-$1034, powering the PWN \citep{j1833_05,j1833_06}. 
It has a spin period $P = 62$\,ms, a spin-down rate $\dot P=2.0\times10^{-13}$,
surface magnetic field strength $B = 3.6\times10^{12}$\,G, and
spin-down luminosity $\dot E =3.3\times10^{37}$\,erg\,s$^{-1}$ \citep{j1833_06}. 

By measuring the expansion speed of \pwn, its age has been found to be 870\,yr \citep{bb08}. 
This makes \pwn\ the second youngest known PWN in the galaxy, only after
Kes 75 \citep{reynolds+18}.
CO and H{\sc i} absorption measurements suggest a distance of $4.7\pm0.4$\,kpc \citep{j1833_06}. 

\begin{figure*}[ht]
    \centering
    \epsscale{1.1}
    \plottwo{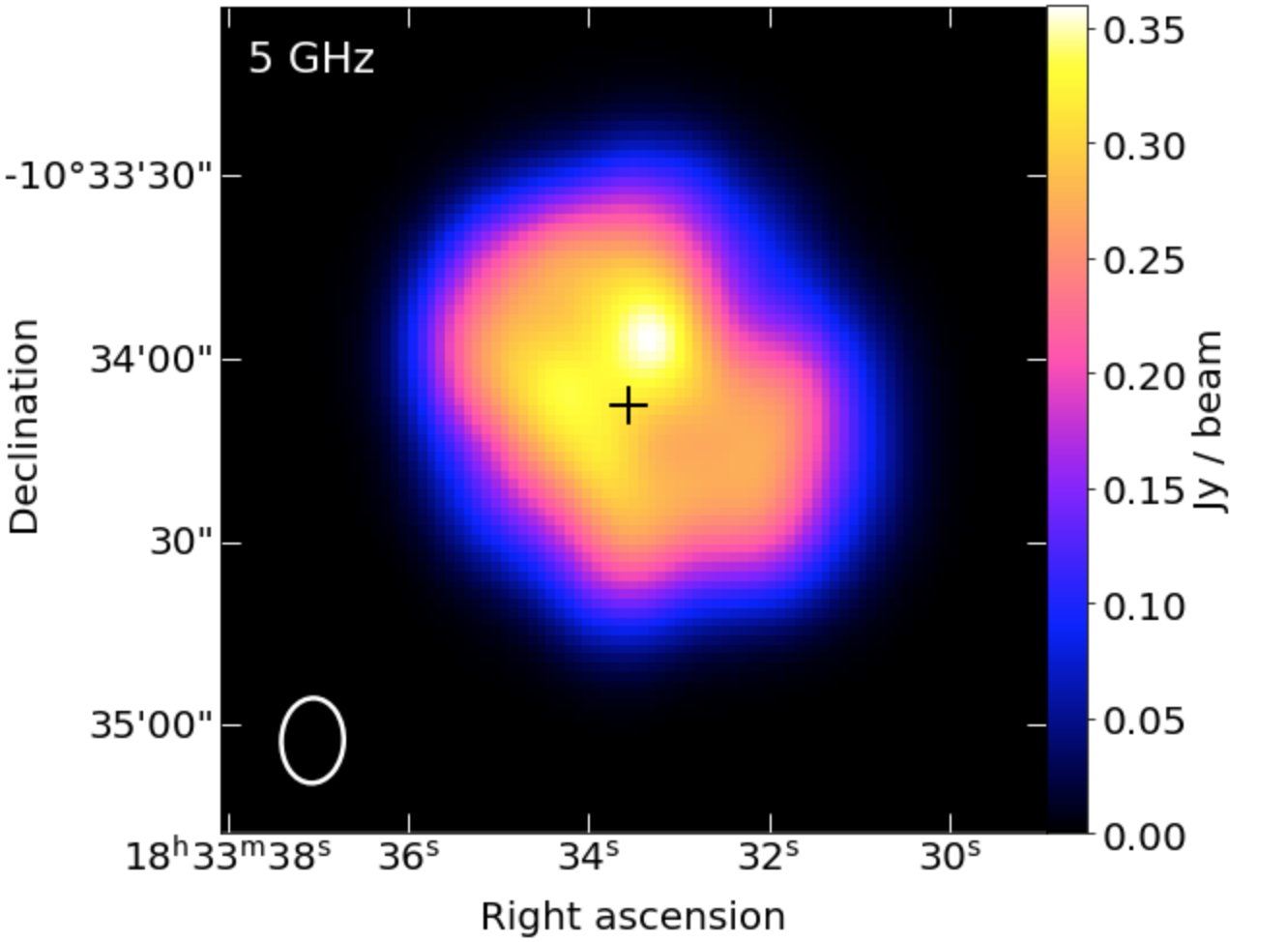}{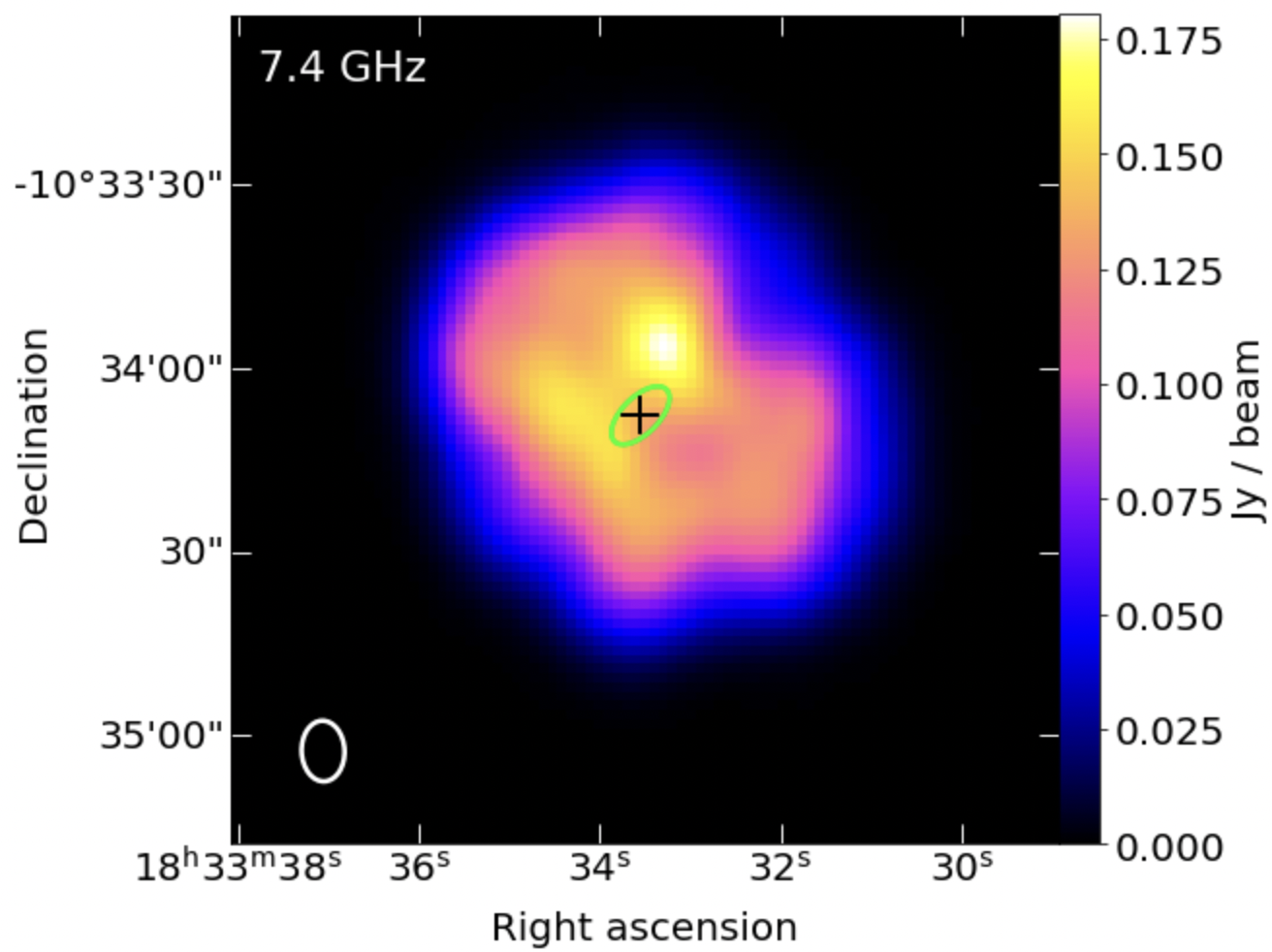}
    \caption{
    Left: total intensity map of \pwn\  at 5\,GHz. The beam size is $13\farcs9 \times 10\farcs2$. The rms noise is 0.09\,mJy\,beam$^{-1}$.
    Right: the same plot at 7.4\,GHz. The position and orientation of the X-ray torus is indicated by the green ellipse \citep[not in scale;][]{ng08}. The beam size is $9\farcs9 \times 7\farcs0$ and the rms noise is 0.07\,mJy\,beam$^{-1}$.
    In both panels, the cross marks the position of PSR J1833$-$1034 and the beam size is indicated at the lower left. 
\label{fig:intensity}}
\end{figure*}

Surrounding the pulsar is compact X-ray emission of radius $\sim$3\arcsec--6\arcsec, which corresponds to the TS
\citep{guest19}, and it has been fit with a torus model to
infer the pulsar spin axis orientation  \citep[][see also Figure~\ref{fig:intensity}]{ng08}.
In radio, a high-resolution Very Large Array (VLA) study revealed the filamentary structure of the PWN \citep{bb08}. 
Different spectral indexes of \pwn\ have been reported at around 5\,GHz,
from $\alpha = -0.08^{+0.09}_{-0.06}$ ($S_\nu \propto \nu^{-\alpha}$) to
$+0.06 \pm 0.03$ to $+0.12 \pm 0.03$  \citep{bb08,archive, sun+11}\footnote{In \citet{archive}, the sign of the spectral index reported in the text is different from that in Figure 1. Through private communication, we have confirmed that the text has a typo and the image shows the correct spectral index.}.



  

The first radio polarimetric observation of \pwn\  was performed over 40 years ago \citep{bs81} and it revealed a radial magnetic field structure. The study, however, did not investigate the detailed distribution of the Faraday rotation in the source. Later polarimetric studies were only done at higher frequencies, for which the Faraday effect is negligible \citep{reich02}. In this study, we derive the rotation measure (RM) toward \pwn\  and map the intrinsic $B$-field orientation at  high resolution, using archival VLA data. The observation details and imaging methods are described in Section~\ref{sec:data} and the results are presented in Section~\ref{sec:obs_results}. We discuss the observed features in Section~\ref{sec:discuss}. In Section~\ref{sec:model}, we show that a turbulent radial magnetic field model is able to reproduce the observed features. We summarize our findings in Section~\ref{sec:conclusion}.

\section{Observations and Data Reduction} 
\label{sec:data}
\begin{figure*}[ht]
    \centering
    \epsscale{1.1}
    \plottwo{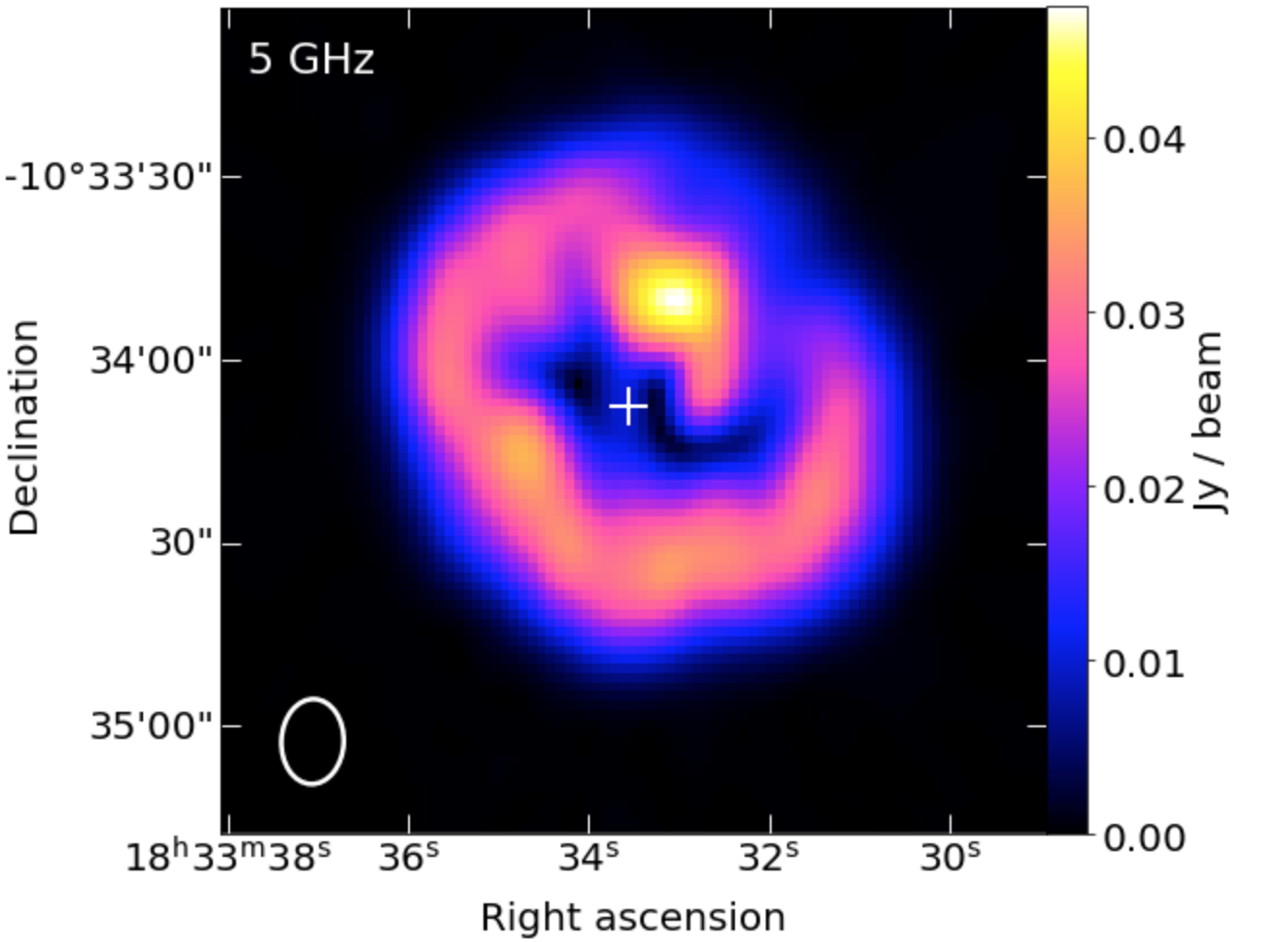}{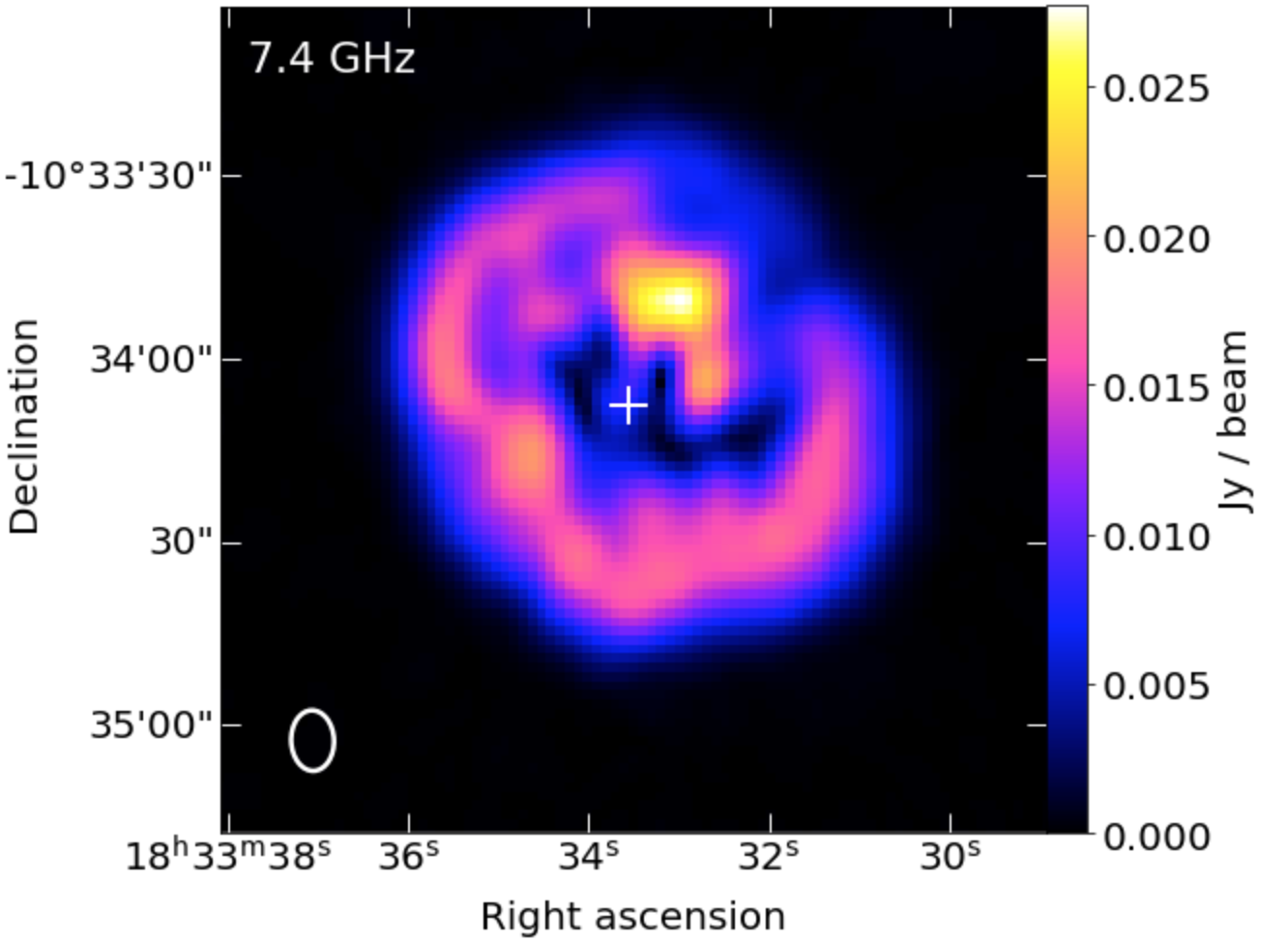}
    \caption{Linear PI maps of \pwn . The crosses mark the position of PSR J1833$-$1034. The beam sizes are same as in Fig.~\ref{fig:intensity} and are indicated at the lower left.
\label{fig:poli}}
\end{figure*} 


We analyzed an archival VLA observation of \pwn\  taken on 2010 August 12 in the C-array configuration. These are the data that have the highest resolution with full Stokes parameters recorded. The data were previously used to study the radio morphology and spectrum of the source, but no polarization property of the source was reported \citep{archive}. The observation parameters are listed in Table~\ref{tab:obs}. The observations were done in the C band, and were divided into two separate frequency bands centering at 5 and 7.4\,GHz, with a bandwidth of 1.024\,GHz each. The observations have $u$-$v$ coverage that is sensitive to scales from $10''$ to $5'$. This well covers the PWN size of $80''$. The data were taken in the continuum mode, with two hours of total on-source time spanning over seven hours. J1331+3030 was observed as a flux and polarization angle (PA) calibrator. J1822$-$0938 was observed at 20 minute intervals to determine the complex gain and leakage terms.

\begin{table}[t!] 
\centering
\caption{Observational Parameters}
\begin{tabular}{p{4.0cm} l}
\hline 
Observing date \dotfill & 2010 August 12 \\
VLA configuration \dotfill & C \\
Number of antennas \dotfill & 23 \\
Baseline \dotfill & 30--1030\,m \\
$u$-$v$ coverage \dotfill & 0.45--19\,k$\lambda$, 0.69--27\,k$\lambda$ \\
Center freq. (GHz) \dotfill & 5, 7.4 \\
Bandwidth (GHz) \dotfill & 1.024, 1.024 \\
Channel width (MHz) \dotfill & 2 \\
On-source time (min) \dotfill & 115 \\
\hline
\end{tabular}
\label{tab:obs}
\end{table} 

We used Common Astronomical Software Application (CASA) version 5.6 for all the data reduction \citep{casa07}. Calibrations were done manually, because the data are too old to run the current pipeline. We first applied the necessary flaggings, including edge channels, shadowed antennas, zero-amplitude data, etc. Then we identified and flagged the radio frequency interference using the task \texttt{rflag}. Next, the flux, bandpass, gain, and polarization calibration solutions were determined using the two calibrators mentioned above. Finally, we employed self-calibration to boost the signal-to-noise ratio (S/N). The dynamic range was nearly doubled after a couple of iterations. We formed Stokes I, Q, and U images, and performed deconvolution using the task \texttt{tclean}. To optimize between sensitivity and resolution, we chose the Briggs weighting algorithm with robust = 0 for all images. 
Since the observations were done in two separate frequency bands, we formed two images at 5 and 7.4\,GHz. This is different from the previous study \citep{archive}, so that we can cross-validate the results. In the final maps, the synthesized beams have FWHMs of $13\farcs9 \times 10\farcs2$ and $9\farcs9 \times 7\farcs0$  at 5\,GHz and 7.4\,GHz, respectively. The rms noises in the Stokes I image are 90 and 70\,$\mu$Jy\,beam$^{-1}$ at 5 and 7.4\,GHz, respectively, and in the Stokes Q and U images are 50 and 45\,$\mu$Jy\,beam$^{-1}$ at 5 and 7.4\,GHz, respectively. The observed rms noise is similar to the theoretical sensitivity.

The Q and U images are used to produce the linear polarization intensity (PI) and PA maps, where ${\rm PI}=\sqrt{Q^2+U^2}$ and ${\rm PA} = \frac{1}{2}\arctan{(U/Q)}$.
We corrected the Ricean bias in the former \citep{ricean} using the measured rms noise levels mentioned above. The observed PAs are in general different from the intrinsic ones because of the Faraday rotation. The change in the PA is proportional to the square of the wavelength as $\Delta{\rm PA} = {\rm RM}\,\lambda^2$, where RM is the rotation measure and $\lambda$ is the observing wavelength.
To solve for the RM, we generated two images at 5 and 7.4\,GHz, each with identical resolutions.
We first applied a Gaussian taper of $14\arcsec\times11\arcsec$ to the visibility data, then followed the same imaging and cleaning procedures as described above.
The final images were convolved from the model images by the same beam size as the taper ($14'' \times 11''$).
The PA and total intensity maps were then used to solve for the RM and spectral index, respectively.
We further divided the data into four frequency bands to confirm that there was no $n\pi$-ambiguity in our case.
We used the RM obtained from the two-band result because it has a better S/N.
The RM is only calculated for pixels of which the PI is larger than 1\,mJy\,beam$^{-1}$ at both frequencies, such that the uncertainty of the RM is smaller than 20\,rad\,m$^{-2}$.
The large S/Ns of the images make the uncertainty in RM that small, despite the small difference in $\lambda^2$.
We also note that the RM is low enough that bandwidth depolarization is negligible. 
Finally, we employ the RM map to derotate the PA map at 7.4\,GHz to obtain the intrinsic $B$-field direction.

\section{Results} 
\label{sec:obs_results} 
Our total intensity maps of \pwn\  at 5 and 7.4\,GHz are shown in Figure \ref{fig:intensity}.
The PWN has an overall spherical shape.
It appears to be more extended at 5\,GHz due to the larger beam size and higher surface brightness.
The pulsar was not detected.
We marked its position in the images using the reported values \citep[R.A., decl.\ $= 18^{\rm h}33^{\rm m}33\fs57, -10\degr34\arcmin07\farcs5$;][]{j1833_06}.
The PWN emission peaks at 12\arcsec\ NW of the pulsar.
The brightest regions form a double-lobed structure, at about equal distance NW and SE from the pulsar.
The double-lobed structure is more obvious at 7.4\,GHz because of the better resolution. 
The two lobes are separated by $\sim16\arcsec$.
The X-ray torus of \pwn\ ($11\farcs4$ in size) appears to be aligned with the lobes (Figure~\ref{fig:intensity}), hinting that they are related.
This feature is similar to that reported in previous studies \citep{fhm+88, bb08}.
Even finer structures of \pwn\ was observed at a higher resolution \citep{bb08}.
We will focus on the polarization results in this paper. 

The integrated flux density of \pwn\  is $6.6\pm0.1$\,Jy and $6.1\pm0.1$\,Jy at 5 and 7.4\,GHz, respectively.
The uncertainties reported above are mostly systematics estimated by using different source regions.
The uncertainties due to calibration and statistical fluctuations are negligible, since the PWN is very bright.
The flux density measurements give a spectral index of $\alpha=0.16 \pm 0.08$ ($S_\nu \propto \nu^{-\alpha}$).
We also tried the spectral tomography method \citep{tomography},
and found a similar value and no significant spatial variation in the spectral index.
Our results are consistent with those reported by \citet{archive}, using a slightly different technique.

\begin{figure}[t!]
    \centering
    \includegraphics[width=0.45\textwidth]{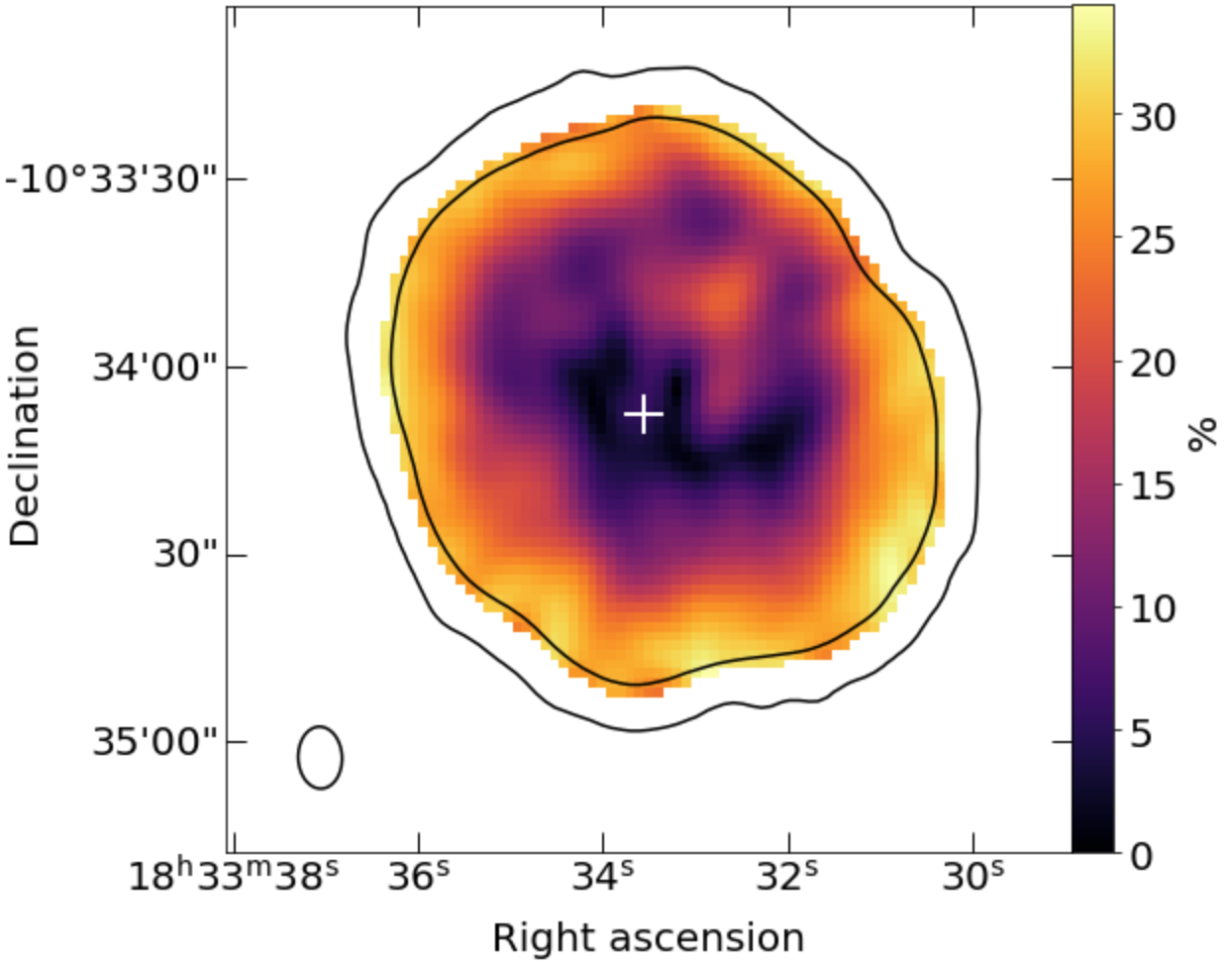}
    \figcaption{Linear PF of \pwn\  at 7.4\,GHz. The map is clipped if the uncertainty is above 1\%. The contours represent the 7.4\,GHz surface brightness at 1 and 10\,mJy\,beam$^{-1}$.
\label{fig:pf5}}
\end{figure}  
\begin{figure}[t!]
    \centering
    \includegraphics[width=0.44\textwidth]{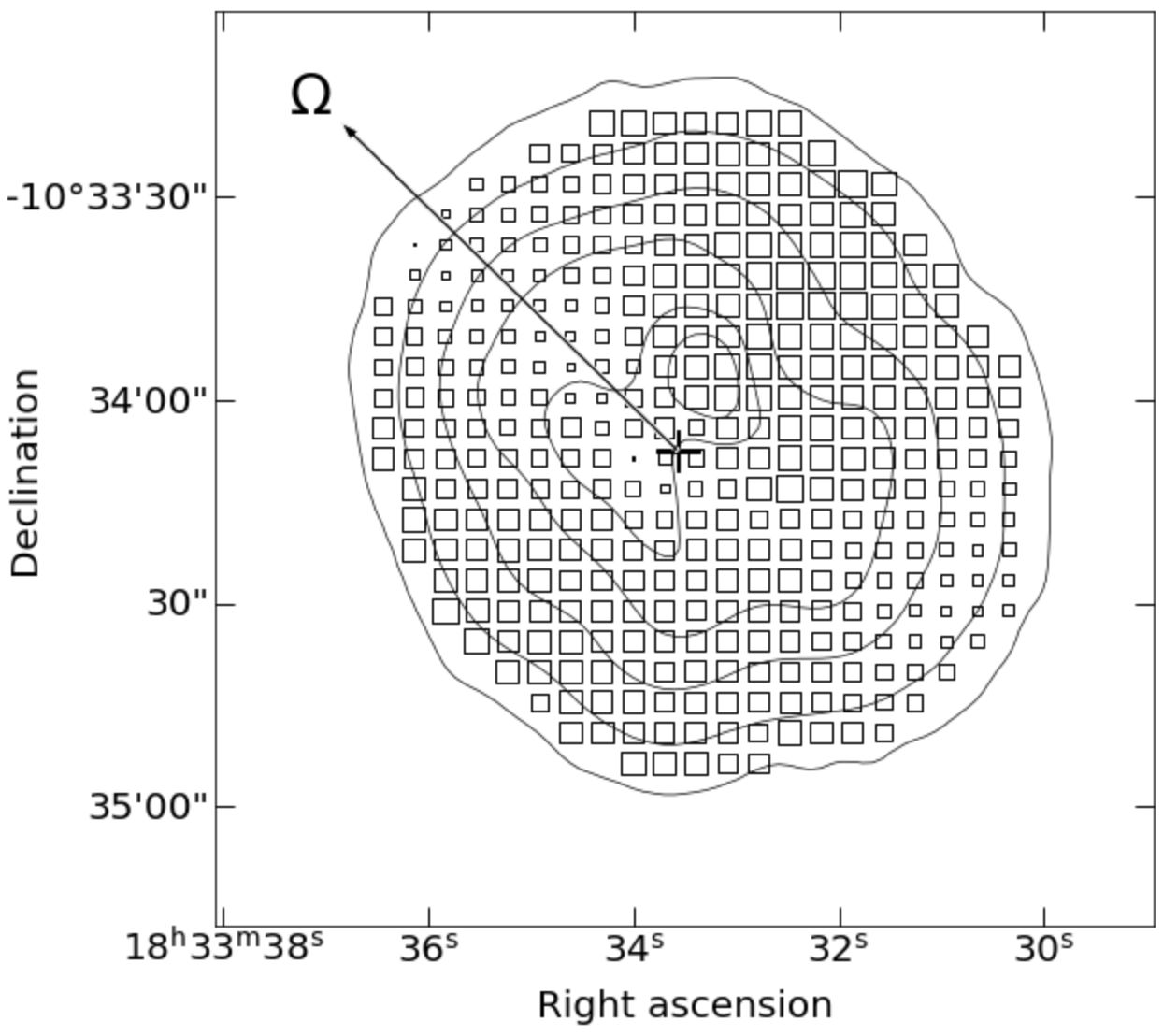}
    \figcaption{RM of \pwn. The boxes scale linearly from +25 to +105\,rad\,m$^{-2}$. The uncertainty is below 20\,rad\,m$^{-2}$. The contours represent the 7.4\,GHz total intensity at 160, 140, 100, 50, 10, and 1\,mJy\,beam$^{-1}$. The arrow indicates the projected spin axis direction of PSR J1833$-$1034.
\label{fig:rm}}
\end{figure}  

\begin{figure}[t!]
\centering
\includegraphics[width=0.455\textwidth]{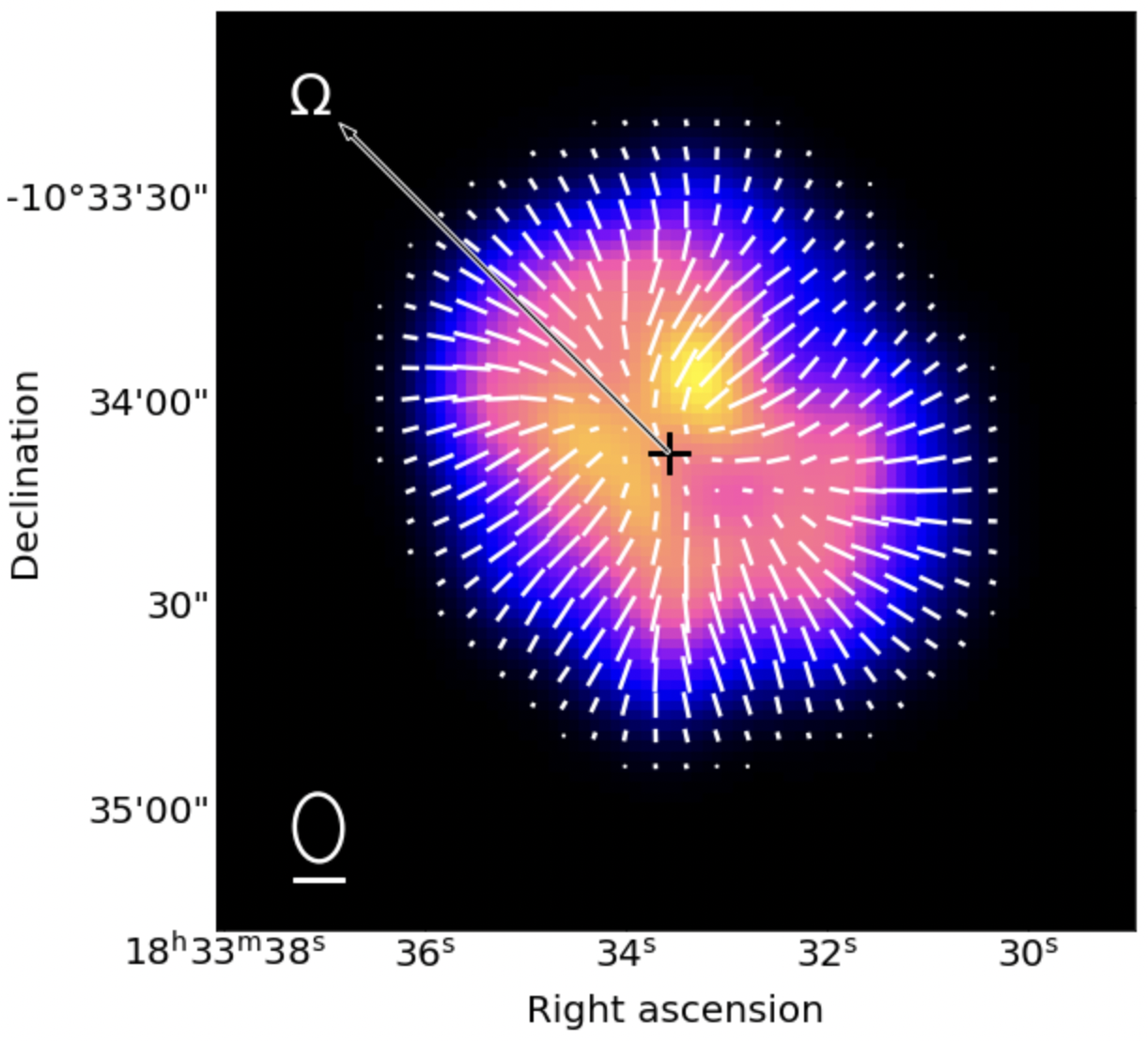} 
\figcaption{Total intensity map of \pwn\  at 7.4\,GHz overlaid with the intrinsic magnetic field orientation. The lengths of the vectors are proportional to the polarized intensity. The black cross indicates the position of PSR J1833$-$1034. The magnetic field map has a beam size of FWHM $9\farcs9 \times 7\farcs0$, as indicated by the ellipse at the lower left. The bar below indicates polarized flux density of 20\,mJy\,beam$^{-1}$. The vectors are clipped if the uncertainty in the PA is larger than $4^\circ$. The arrow shows the projected spin axis direction of PSR J1833$-$1034 \citep{ng08}. 
\label{fig:mag}}
\end{figure}

\begin{deluxetable*}{ccccc}
\centering
\caption{Double-lobed Radio PWNe \label{tab:lobes}}
\tablehead{
 & \pwn\  & Vela  & G343.1$-$2.3 & G76.9+1.0
}
\startdata
         Age (kyr) \dotfill & 0.87 & 11& 18 & 9$\sim$40  \\
         Separation between two lobes (pc) ($l_{\rm lobes}$) \dotfill & 0.36 & 0.36 & 1.24  & 8.7 \\
         Peak brightness ratio between two lobes \dotfill & 1.07 & 2  & 1.14 & 1.08\\
         Size of X-ray torus (pc) ($l_{\rm torus})$ \dotfill & 0.26  & 0.06 & 0.1 & \nodata \\
         $l_{\rm lobes}/l_{\rm torus}$ \dotfill & 1.4 & 6 & 12 & \nodata \\
         $\dot{E}_{\rm pulsar} \, (10^{36} \, {\rm erg\,s}^{-1})$ \dotfill & 34 & 6.9   & 3.4& 30\\
\enddata
\tablerefs{The X-ray torus sizes are from \citet{ng08}.}
\end{deluxetable*}

Figure \ref{fig:poli} shows the PI map of the PWN. The PI is low at the center and it peaks near the NW lobe, whereas the polarized emission near the SE lobe is not obvious.
The position of the peak of the PI is offset by around 7\arcsec\ from the peak of the total intensity.
Figure \ref{fig:pf5} shows the linear polarization fraction map (PF = PI/I) of \pwn.
The overall PF is $\sim 13\%$ at both 5 and 7.4\,GHz. 
The PF mainly depends on the magnetic field structure and magnetic turbulence; thus, this map provides different information from the PI map.
The PF increases gradually from about zero at the center to $\sim$30\% near the boundary.


Figure \ref{fig:rm} shows the RM map derived from the PA maps at the two bands. The RM values range from +25 to +105\,rad\,m$^{-2}$ with uncertainty less than 20\,rad\,m$^{-2}$.
The RM of the nebula near the pulsar position is about +50\,rad\,m$^{-2}$, which is similar to the +60\,rad\,m$^{-2}$ measured from the pulsed emission of PSR J1833$-$1034 \citep{serylak+21}. 
Since the pulsar emission only passes through half of the nebula, its RM value can be different from that of the PWN if the internal Faraday rotation is strong.
In Figure \ref{fig:rm}, we illustrate the spin axis direction of the central pulsar inferred from the X-ray torus \citep{ng08}. The RM map shows a symmetric pattern that aligns with the pulsar spin direction. The RM values are lower along the axis and gradually increase outward.
The symmetry can also be seen in the intensity map,
but not in the PI map.

We use the RM map to correct for the Faraday rotation.
The resulting intrinsic orientation of the  $B$-field vectors (PA$+90^\circ$) is plotted in Figure~\ref{fig:mag}.
The PA uncertainties in the map are smaller than $4^\circ$.
We found an overall radial field, as reported in previous works \citep{bs81,reich02}.
In addition, our high-resolution map shows that some magnetic field vectors appear to deviate from the radial direction by at most $45^\circ$. The deviation is most evident at the bright double lobes, forming an hourglass-like structure with $\sim$30\arcsec\ across. Observations with better resolution and at higher sensitivity are needed to confirm this feature.


\section{Discussion} \label{sec:discuss}

\subsection{Double-lobed Structure}
The double-lobed structure in \pwn\  is also observed in other radio PWNe, including Vela \citep{dlmd03}, G343.1$-$2.3 
\citep{romani_ng+05, yihan+22}, 
and G76.9+1.0 \citep{landecker+93, arz+11, aru+14}. 
3C58 \citep{reich02, bie06} and DA 495 \citep{kothes+08} show only a double-lobed structure in the polarized emission, but not in the total intensity.
The positions of the lobes are always symmetric about the pulsar spin axis, except for DA 495 whose pulsar spin 
orientation is unknown.
The symmetry suggests that the X-ray torus might be related to the lobes.
The observed double-lobed morphology of the Vela and G76.9+1.0 PWNe was suggested to result from the toroidal emission region, with specific viewing geometry \citep{Chevalier+11}. This model, however, may not be applicable to the spherical PWN like G21.5$-$0.9.

All these lobes are only seen in radio, suggesting the accumulation of less energetic particles in these regions. 
It is peculiar that they all show a double-lobed morphology, but not a ring-like one, as in the X-ray torus.
In Table~\ref{tab:lobes} we summarize the properties of four double-lobed radio PWNe. 
We notice that the two lobes always show different brightness.
Vela has the largest peak brightness difference between the two lobes.
There is also a large range in the physical separation of the lobes, 
from 0.33\,pc in \pwn\  to 8.7\,pc in G76.9+1.0.
From the table, there seems to be a trend that older PWNe have larger lobe separation.
This trend also exists for the ratio of the separation to the size of the X-ray torus.
If confirmed, this could imply an outward motion of the lobes.



\subsection{Internal Faraday Rotation} \label{sec:dis:ifr}
\begin{figure}[t!]
    \centering
    \includegraphics[width=0.47\textwidth]{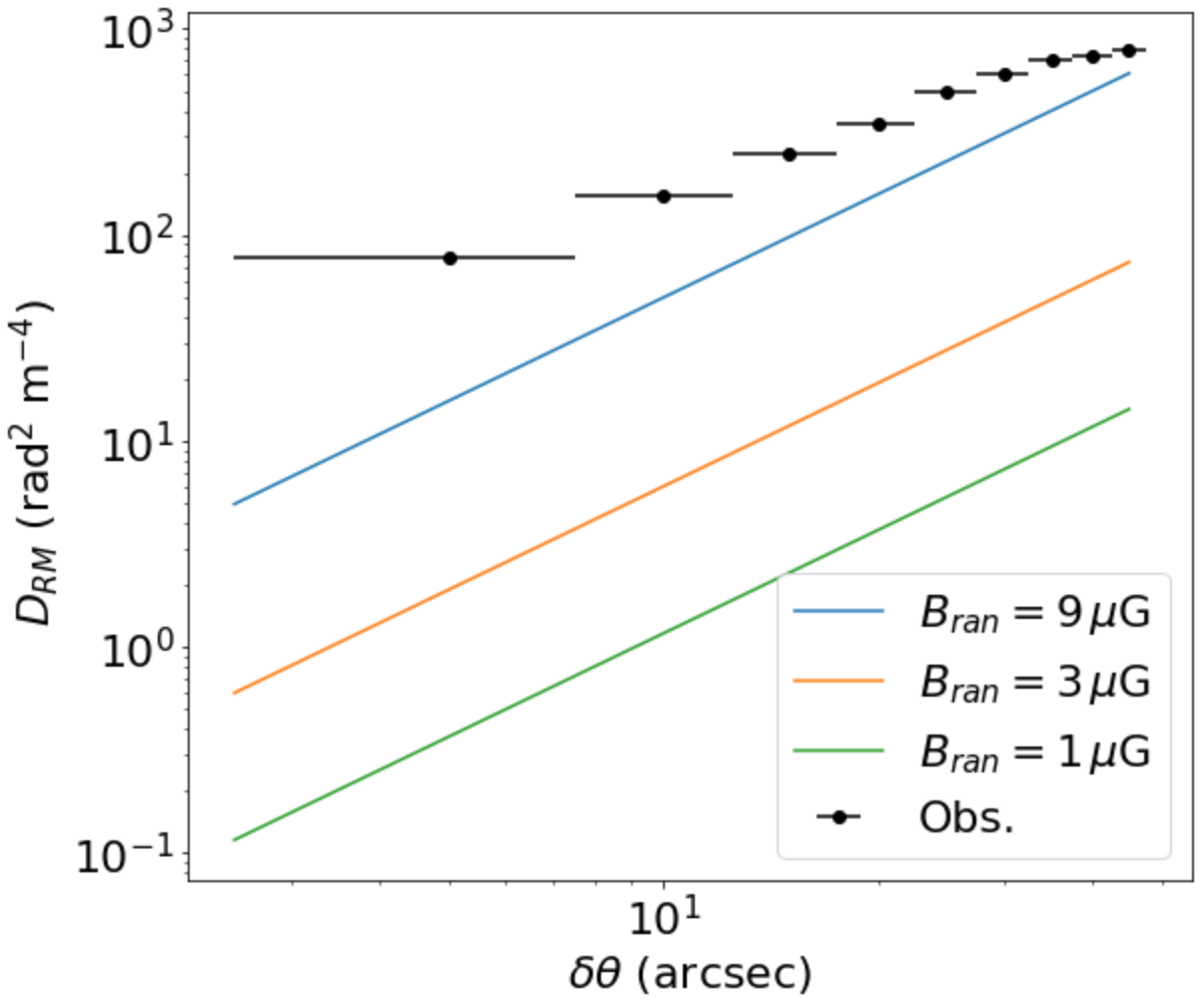}
    \figcaption{Structure functions of the RM for \pwn. The black dots are the observational results calculated from the RM map in Figure \ref{fig:rm}. The colored lines show the theoretical model with different strengths of the magnetic fluctuation \citep{minter+96}.
    \label{fig:structure_func}}
\end{figure}

It is intriguing that the RM toward \pwn\ appears to show a symmetric pattern about the pulsar spin axis (Figure \ref{fig:rm}).
Instead of the chance alignment of the foreground RM structure,
this more likely results from the Faraday rotation within the PWN.
This is supported by the amplitude of the RM variation.
\citet{minter+96} derived a structure function for the foreground RM,
$D_{\rm RM}(\delta\theta)$, in terms of angular scale,
due to the fluctuation of both the interstellar electron density and the galactic magnetic field caused by turbulence
(see the Appendix \ref{sec:appen} for details).
The theoretical structure function predicts the average RM difference for a given angular separation.
For the case of G21.5$-$0.9, we obtain $D_{\rm RM}(\delta\theta) \approx (13B^2_{\rm ran}+11)\delta\theta^{5/3}$\,rad$^2$\,m$^{-4}$, where $B_{\rm ran}$ is
the strength of the random $B$-field component in units of $\mu$G and $\delta \theta$ is the angular separation in units of arcmin. 
The value of $B_{\rm ran}$ depends on the observation direction, and its largest empirical value is 9\,$\mu$G \citep{haverkorn+08}. This model can be compared with our observations, and the result is shown in Figure \ref{fig:structure_func}. The structure function is calculated up to an angular scale of $45''$, the same as the PWN size. The observed structure function is larger than the theoretical values at all $\delta\theta$ for $B_{\rm ran} \leq 9\,\mu$G.
This suggests that the foreground Faraday rotation is unable to explain the rapid variation of RM in \pwn\  unless $B_{\rm ran}$ is exceptionally high.
This result indicates that the observed RM variation is unlikely to be due to the foreground.

Pulsar wind in general cannot contribute significant Faraday rotation.
The main reason is that it consists of electron-positron plasma, such that the Faraday effect is canceled out. 
Moreover, Faraday rotation is weak for relativistic particles \citep{jones_odell77}, and
their number density in pulsar wind is low; for example, less than $10^{-5}\,$cm$^{-3}$ for \pwn\ \citep{hattori+20}. 
On the other hand, the RM fluctuation we observed 
($\sim$80\,rad\,m$^{-2}$) can be explained by electrons originating from the SN ejecta. 
%
For a magnetic field strength of $130\,\mu$G \citep{guest19} and a nebula size of 2\,pc,
the observed RM value requires an average electron density of $\sim$0.8\,cm$^{-3}$.
We have doubled the density to account for the effect of the internal Faraday rotation \citep{burn66}.
Assuming that \pwn\ is a sphere, the total number of cold electrons can be
contributed by 0.1\,$M_\sun$ of hydrogen or 0.2\,$M_\sun$ of high-Z elements.
This is possible for \pwn\  since the swept-up ejecta mass is suggested to be 0.8\,M$_\sun$ \citep{hattori+20}.

In this picture, the spatial variation of the RM is then caused by the variation of either the line-of-sight magnetic field strength or ejecta electron density.
The radial $B$-field of \pwn\ is expected to produce a spherically symmetric RM pattern rather than an axisymmetric one. 
Also, nonuniform field strength should result in a correlation between the RM and the surface brightness, which is not observed. 
We therefore conclude that there is an axisymmetric distribution of ejecta electrons about the pulsar spin axis.
This could be due to an asymmetric SN explosion, or a latitude-dependent mixing process between the pulsar wind and the SN ejecta.
Further simulations are needed to understand the details of these processes.

The RM distributions of the Boomerang PWN \citep{kothes06} and CTB 87 \citep{kothes+20, ctb87_xray} show similar patterns as \pwn. They are both symmetric about the pulsar spin axis, and the RM values are lower or more negative along the axis. They have also been suggested to be caused by the internal Faraday rotation \citep{kothes06, kothes+20}. Unlike \pwn, both of them are evolved PWNe that have been crushed by the SN reverse shock. This could imply that the axisymmetric RM pattern already existed before the crush and was not disturbed by the crush. 
However, we note that these two sources have different $B$-field structures than \pwn. The origins of their RM patterns thus may not be the same as for our case.

\subsection{Polarization Properties\label{sec:pl_prop}}

Our high-resolution polarization map confirms that \pwn\ has a radial magnetic field overall.
A previous infrared study found a toroidal field in the inner PWN with scale of 4\arcsec\ \citep[i.e., 0.1\,pc;][]{zajczyk+12}.
If this extends to larger radii, 
its influence on the inner PWN may explain the hourglass structure seen in the PL map (see Section~\ref{sec:obs_results}).
It is, however, unclear what is the extent of the toroidal field is and how it connects to the global radial field of the radio PWN.
Nevertheless, as we will show in Section~\ref{sec:model} below,
we can reject the existence of any large-scale toroidal 
field components with comparable strength.
This suggests that the toroidal field is only important to the inner nebula.


In addition to \pwn, the Crab Nebula and DA~495 also show radial magnetic field structures.
For the Crab, the radial field is only found near the boundary, and it could be caused by fluid instability between the pulsar wind and the ambient medium \citep{crab_mag}.
While numerical simulations suggest that the instability layer can be as thick as 30\% of the PWN radius \citep{bucciantini+04},
this is still insufficient to explain our observations.
The radial field of DA~495 has been suggested to be a combination of dipole and toroidal fields \citep{kothes+08}, which is not the case for \pwn\ (see Section~\ref{sec:model} below).

Finally, we note that as shown in Figure~\ref{fig:pf5}, the PF is very low near the center and gradually increases toward the edge.
This cannot be solely due to beam depolarization, since this effect becomes negligible for angular scales beyond the beam size.
As an example, consider a radial $B$-field emitting at 100\% PF locally; 
the PF is zero at the center, because of beam depolarization, but rises to 85\% just one beamwidth outside.
In Section~\ref{sec:model}, we suggest that the radially increasing PF is caused by the magnetic field geometry and turbulence.
The low PF near the center could also lead to the observed offset between the peaks of the total intensity emission and the polarized emission.

\section{Magnetic Field Models} \label{sec:model}

In this section, we aim to construct a simple model to fit the observed total intensity, PF, and magnetic field orientation of \pwn,
so that we can determine the global $B$-field configuration.
We will then employ the best-fit model to study the effects of internal Faraday rotation.

\subsection{Models Descriptions}


Three-dimensional (3D) simulations have shown that the magnetic field in a PWN could have significant turbulence \citep{porth+14}.
In our model, we consider ordered and disordered (i.e.\ turbulent) $B$-field components, and adopt the method of \citet{ban+pet16} to calculate synchrotron emissivity
in the presence of small-scale magnetic turbulence.
On top of the ordered magnetic field $\bm{B}$,
a Gaussian random field with variance $(|\bm{B}|\sigma)^2$ in each direction is introduced to account for a turbulent environment. The energy ratio between the ordered and disordered fields is thus $B^2_{\rm ordered}/B^2_{\rm disordered} = 1/3\sigma^2$. A larger $\sigma$ would decrease the local PF. 
For simplicity, we assume a constant $\sigma = 0.7$ over the entire nebula.
The value was chosen so that the PF values are similar to the observations.
This gives $B_{\rm ordered}/B_{\rm disordered} \sim 0.8$, which is close to the previous estimate \citep{fhm+88}. 

We considered three models for the ordered magnetic field.
The first one is a radial $B$-field in 3D (the R model) motivated by our polarization map.
The second one is a toroidal field plus a radial field (the TR model) because the former was observed in the inner PWN \citep{zajczyk+12}.
The third model is a toroidal field plus a dipole field (the TD model),
which was used to explain the observations of DA 495 \citep{kothes+08}.
Mathematically, these can be expressed in spherical coordinates $(r, \theta, \phi)$ as 
\begin{equation}
    \bm{B_R}(r) = -B_0 \hat{r}, 
\end{equation}
\begin{equation}
    \bm{B_{TR}}(r, \theta) = B_0 (\sin\theta \hat{\phi}-\hat{r}), 
\end{equation}
and 
\begin{equation}
    \bm{B_{TD}}(r, \theta) = B_0 (\cos{\theta} \hat{r} + \frac{1}{2} \sin{\theta} \hat{\theta} + \sin\theta \hat{\phi}),  
\end{equation}
where $B_0$ is a constant to characterize the field strength.
The form of the toroidal component in the TR and TD models is adopted from \citet{dva+06}. 
For simplicity, we consider equal strengths for the radial, toroidal, and dipolar field components.
The pulsar spin axis points to the $+z$ direction, and $\theta$ is the polar angle measured from the spin axis.



We note that the directions of the radial field (inward or outward) and the toroidal field (clockwise or counterclockwise) is not constrained by the surface brightness map
because they give the same synchrotron emissivity.
However, they induce different signs of internal Faraday rotation.
An outward radial field produces a positive RM and an inward field produces a negative RM.
The average RM of \pwn\ is $\sim$65\,rad\,m$^{-2}$ near the edge and $\sim$50\,rad\,m$^{-2}$ at the center.
This suggests a negative RM contribution by the nebula, hence an inward radial field.
The direction of the toroidal field is randomly picked.
Except for the magnetic field configuration, the other settings of the models are the same.

The viewing angle between the pulsar spin axis and the line of sight for \pwn\ is 85.4\arcdeg\ \citep{ng08}.
We take it as 90\arcdeg\ in the modeling.
We consider only the region from the TS ($R_{\rm TS} = 0.13\,$pc) to the outer radius ($R_{\rm pwn} = 1.14\,$pc)
and assume no emission outside this region.



To compute the synchrotron emissivity, we also need to specify the pulsar wind particle distribution.
We adopt the diffusion model deduced from the X-ray photon index and TeV brightness profiles \citep{tang+cheva12,Tev17}.
For injected particles following a power-law energy distribution, $N(E, r=0) = KE^{-(2\alpha+1)}$,
there is an exact formula for the density profile
\citep[see Eq.~17 in][]{gratton72}.
When $E \ll 422/(B_0^2t)$ and $r \ll \sqrt{Dt}$ (in cgs units), this can be approximated by
\begin{equation} \label{eq:density}
    n_{pwn}(r) \propto \frac{K}{Dr}  \quad \mbox{for } R_{TS} < r < R_{pwn},
\end{equation}
where $D$ is the diffusion coefficient and $t$ is the age of the PWN.
For our case, the first condition is easily satisfied, given that we are dealing with radio-emitting electrons. The diffusion coefficient $D$ is found to be $3.7\times10^{27}\,{\rm cm}^2\,{\rm s}^{-1}$ by fitting the X-ray data of \pwn\  \citep{tang+cheva12}.
Although $D$ could be smaller for less relativistic particles, the difference is subtle. Given that the PWN age is 870\,yr and the radius is around 1\,pc, the second limit is approximately fulfilled.
We do not attempt to fit $B_0$ and $K$, but arbitrarily set their values and only focus on the relative brightness. The spectral index $\alpha$ is taken to be $+0.16$ derived from our observations.

\begin{figure*}[t!]
    \centering
    \includegraphics[width=0.99\textwidth]{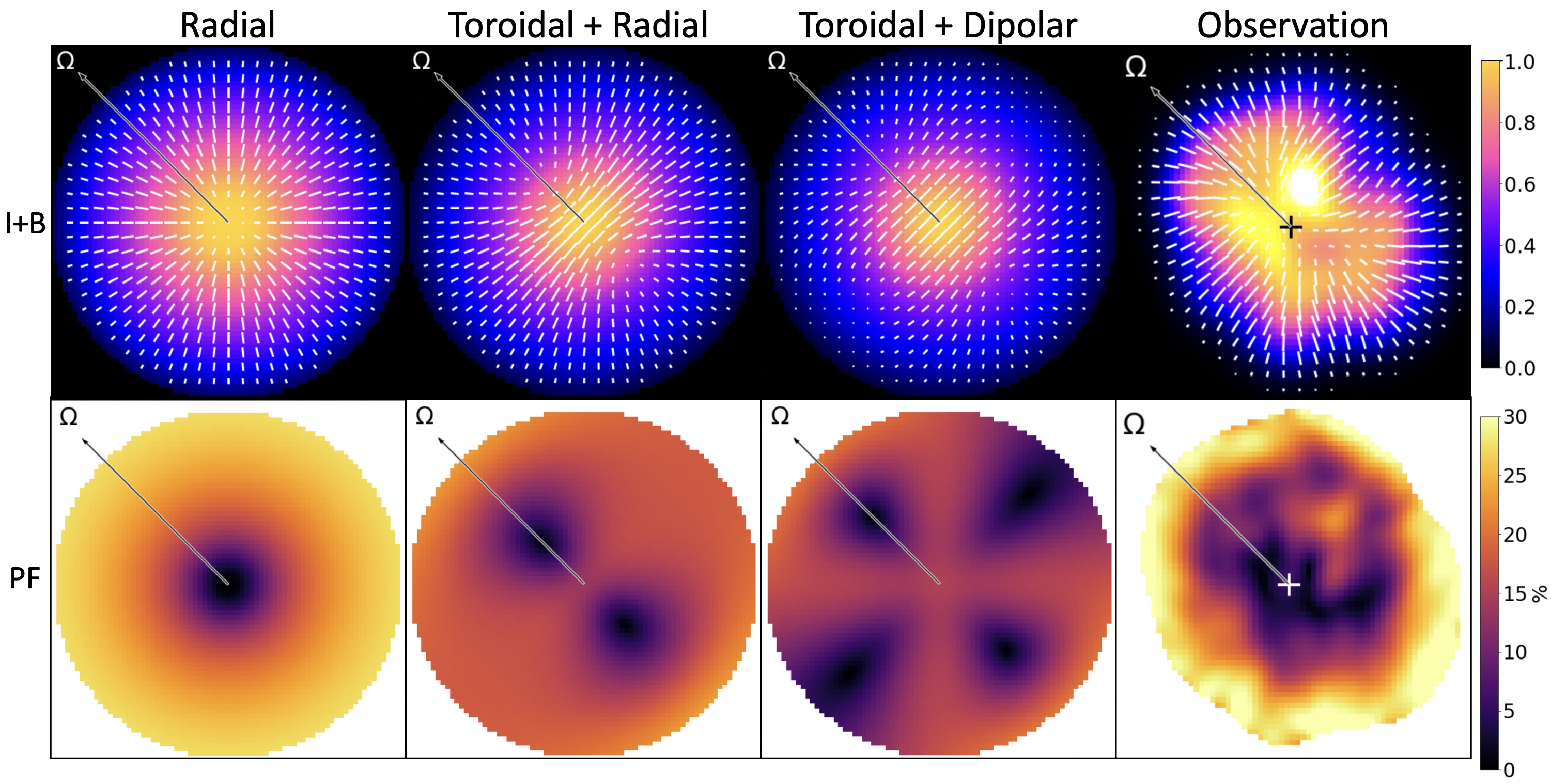}
    \figcaption{Simulated results of the radial field (R) model, the toroidal plus radial field (TR) model, and the toroidal plus dipolar field (TD) model. The bservation results are shown on the right for comparison. The upper row shows the normalized surface brightness overlaid with the magnetic field orientation. The lower row shows the PF maps.
\label{fig:model}}
\end{figure*} 

\begin{figure}[t!]
    \centering
    \includegraphics[width=0.47\textwidth]{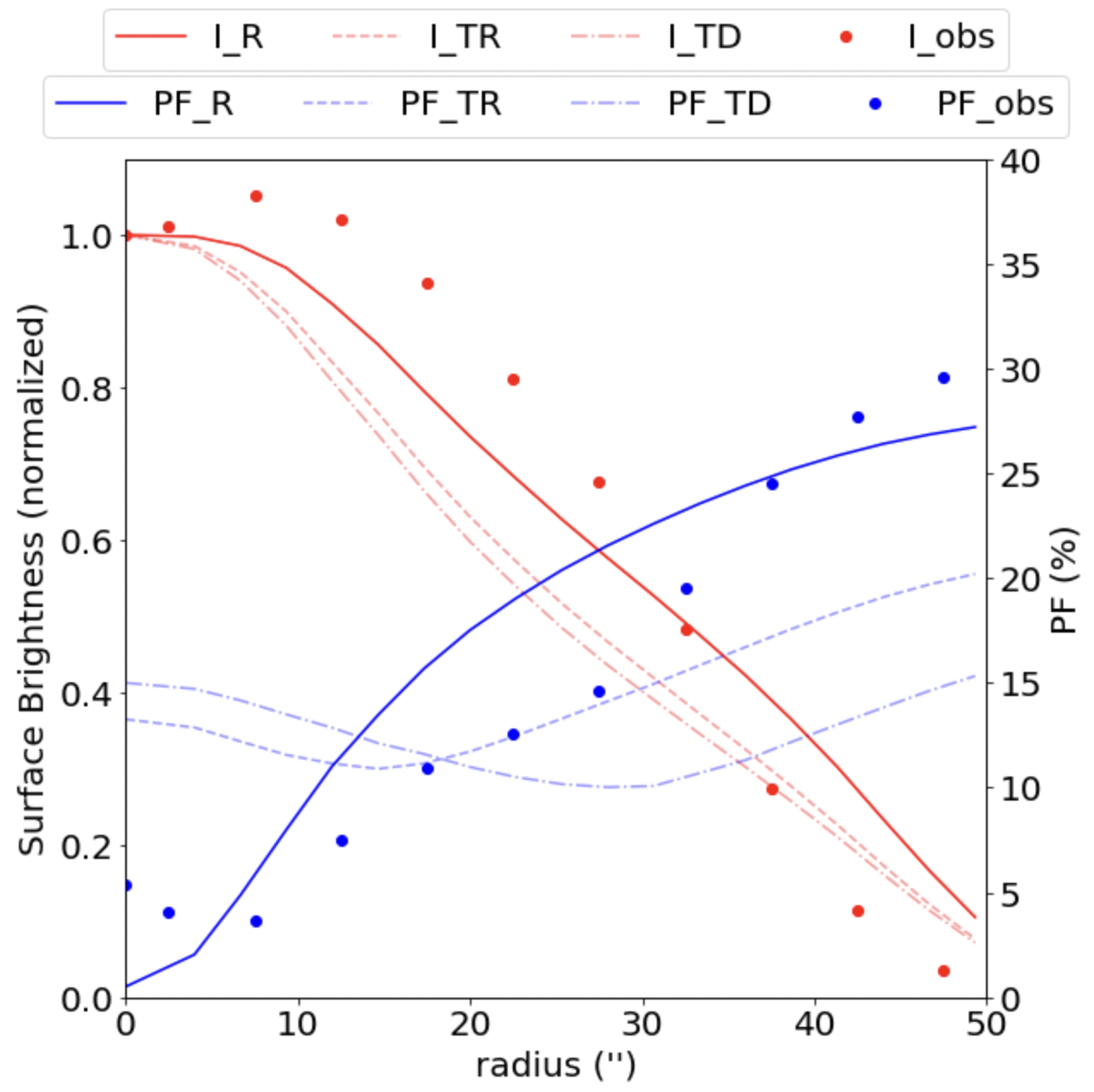}
    \figcaption{Observed and simulated radial surface brightness and PF profiles of \pwn. The red curves and dots are the normalized surface brightness. The blue curves and dots are the PF. The solid lines, dashed lines, dashed-dotted lines and dots represent the R model, the TR model, the TD model, and the observations respectively.
\label{fig:radial}}
\end{figure} 

We followed the recipe derived by \citet{ban+pet16} for deriving synchrotron emission in the Stokes I, Q, and U parameters.
The final maps in each Stokes parameter are obtained by integrating the emission along each line of sight.
We simulate the emission at 5\,GHz, the same as the observations.
Absorption and scattering are negligible at this frequency.
Finally, we smooth the synthetic emission maps by the same beam size as the observations to allow a direct comparison.
As we will show in Section~\ref{sec:model:ifr} below,
depolarization caused by internal Faraday rotation is negligible at this frequency.
We therefore did not account for this effect in our modeling.

\begin{figure*}[t!]
    \centering
    \includegraphics[width=0.99\textwidth]{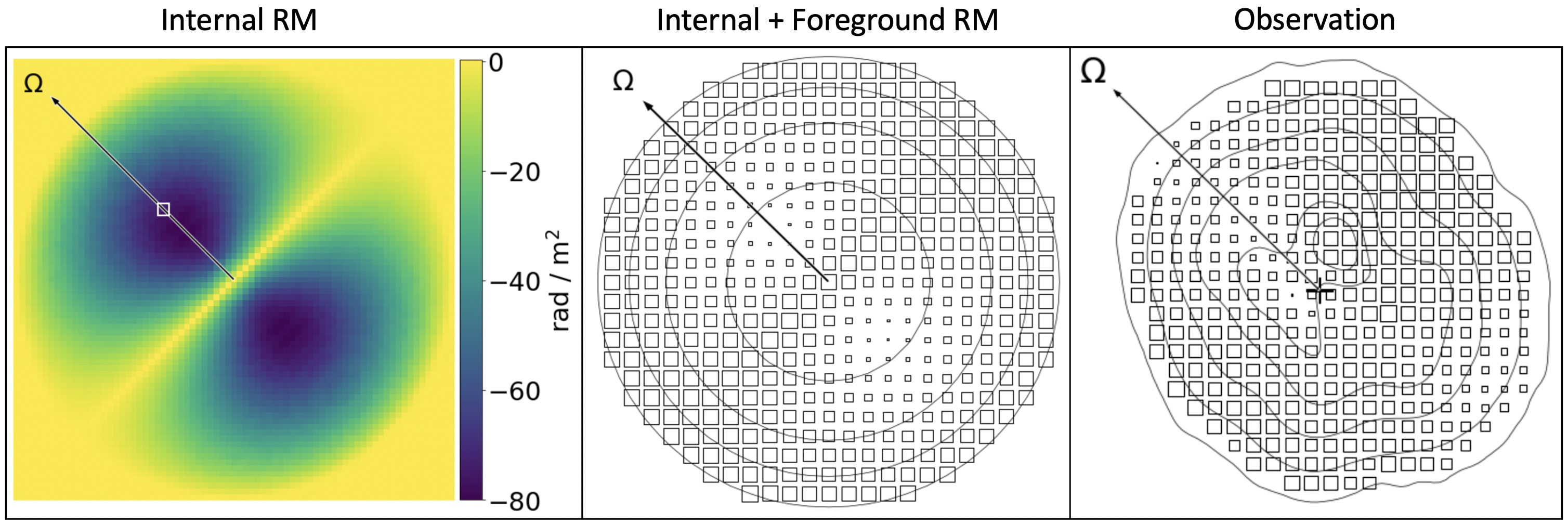}
    \figcaption{Simulated RM maps at 5\,GHz compared with observations. Left: the RM due to internal Faraday rotation. The white box indicates the pixel we used to analyze how the observed values may change with wavelength (see text and Figure~\ref{fig:papf}). Middle: RM map with a foreground RM of 105\,rad\,m$^{-2}$ on top of the internal RM. The scale of the boxes is identical to that for the observations (right panel). 
    \label{fig:rms}}
\end{figure*}

\begin{figure}[t!]
    \centering
    \includegraphics[width=0.47\textwidth]{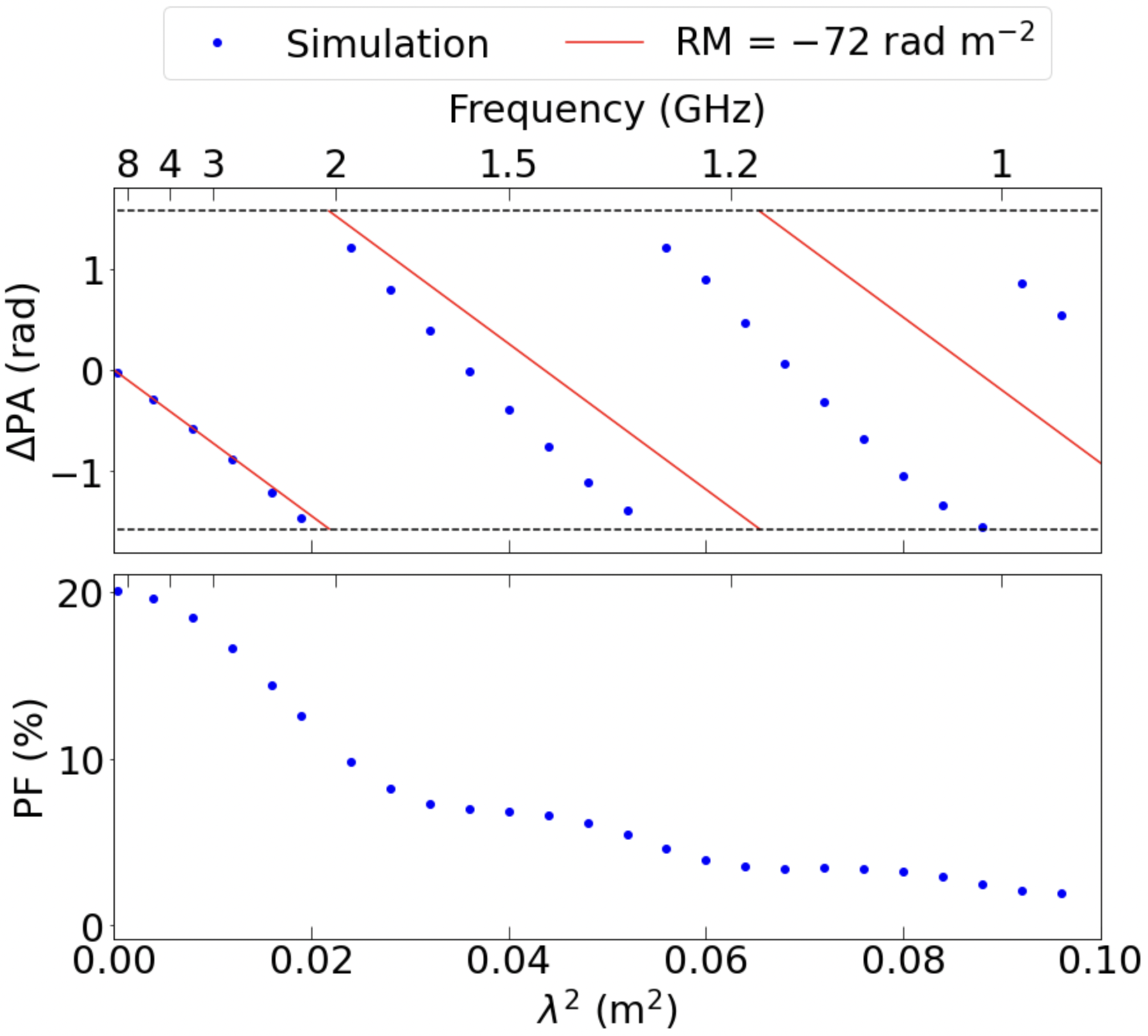}
    \caption{Illustration of the PA change and PF with respect to wavelength for the pixel indicated in Figure~\ref{fig:rms}.
    The blue dots show the simulated results and the red lines indicate a single RM value of $-72\,$rad\,m$^{-2}$ obtained by fitting the PA at short wavelengths (top panel). \label{fig:papf}}
\end{figure}

\subsection{Synthetic Maps}

Figure \ref{fig:model} shows the simulated surface brightness and polarization maps of different magnetic field configurations.
When comparing with the observations, we focus on two prominent features:
the radial magnetic field structure and the radially increasing PF.
The TR model results in an overall radial field structure.
However, there are two local minima in the PF map along the polar axis, rather than a minimum at the center.
The low-PF regions are where the toroidal and radial fields are perpendicular to each other.
The TD model barely reproduces a radial field structure, and the PF map even has four local minima, for the same reason as above.
The R model successfully produces a radial field pattern and the radially increasing PF.

In Figure~\ref{fig:radial} we compare the simulated flux density and PF profiles with the observations.
It shows that the relative surface brightness is insensitive to the magnetic field structure.
The general downward trend is caused by the spherical geometry and the $1/r$ dependence of the pulsar wind particle density.
The differences between the $B$-field configurations are not significant, and all are roughly consistent with the observations.
Note that our model results are very similar to the one with constant particle density and the magnetic field strength decreasing as $1/r$ \citep{fhm+88}.
This is due to the degeneracy of these two parameters for synchrotron radiation.

Unlike the brightness maps, different magnetic field configurations give very different PF maps.
Only the R model shows an increasing trend with radius that generally matches the observations.
We argued in Section~\ref{sec:pl_prop} that the angular size of this trend is too large to be attributed to beam depolarization.
The increasing trend can be understood as follows.
First, considering the case without turbulence (i.e.\ $\sigma = 0$),
if we ignore the Faraday effect, the observed polarization direction is perpendicular to the orientation of the $B$-field projected onto the plane of the sky.
For the R model, picking any line of sight, the projected $B$-field always has the same position angle.
Therefore, after projection, the polarized emission from all cells in 3D adds up coherently, such that the observed PF is a constant over the entire nebula and equal to the intrinsic PF of each cell.
Note that this is not the case for the TR and TD models.
They are expected to have lower PFs in the inner nebula, due to the changing magnetic field PA along the line of sight.

Now, considering turbulence, an analogous formula of the local PF in each cell can be written as 
$PF_{\rm local} \sim \frac{PI\sin\eta}{I(\sin\eta + \sigma)}$,
where $\eta$ is the angle between the line of sight and the magnetic field in 3D.
A large turbulence level $\sigma$ thus leads to a lower PF, and
the latter also depends on $\eta$.
In the R model, the edge of the nebula has $\eta\approx90\arcdeg$, and hence the highest PF, while the PF is lowest near the center as $\eta\approx 0$.
The same result can be derived rigorously using a full mathematical treatment with the synchrotron recipe \citep{ban+pet16}.

We conclude that the R model is sufficient for explaining explain the observed radial profiles of brightness and PF.
This rules out the existence of a global toroidal or dipolar $B$-field components with comparable strength to the radial field.


\subsection{Internal Faraday Rotation} \label{sec:model:ifr}

In the following discussion, we focus on the R model to discuss its effect on the internal RM and depolarization.
We model the ejecta electron density distribution as $n_{sn}(\theta) = n_{sn,0}|\cos\theta|$, since it fits our observational results,
where $n_{sn,0}$ is a constant and $\theta$ is the polar angle measured from the spin axis of the pulsar.
These electrons are cold, such that their contribution to the synchrotron emission is negligible.
We find that for $B_0 = 130\,\mu$G \citep{guest19},
$n_{sn,0}=4.2$\,cm$^{-3}$ can give an RM fluctuation of
$\Delta \mathrm{RM=80}$\,rad\,m$^{-2}$ within the PWN.
This implies a volume-averaged number density of $\sim0.5n_{sn, 0} = 2.1\,{\rm cm}^{-3}$.

Figure~\ref{fig:rms} shows the simulated RM maps.
The internal RM values are all nonpositive because the magnetic field in the near hemisphere points away from the observer.
There is no Faraday rotation at the equator, since we assumed $n_{sn}(\theta = \pi/2) = 0$ in the model. 
For a better comparison with the data, we added a constant foreground RM of 105\,rad\,m$^{-2}$.
The resulting RM map is shown in the middle panel of Figure~\ref{fig:rms}.
The RM values are smaller along the pulsar spin axis,
which generally matches with the observations.
At the PWN boundary, our model suggests minimal internal Faraday rotation due to the short light path.
This gives nearly constant RM value around the boundary,
in contrast to what we have observed.
The discrepancy could be an observational bias, since some pixels near the edge are clipped in the map due to the low surface brightness.

Internal Faraday rotation can lead to depolarization and break the linear relation between PA and $\lambda^2$ at low frequency \citep{burn66}. 
This has been suggested to be the cause of the frequency-dependent depolarization found in the Boomerang and CTB 87 PWN \citep{kothes06,kothes+20}.
We illustrate this effect with a line of sight near the strongest internal RM (see the left panel of Figure~\ref{fig:rms}).
We calculated the Faraday rotation of the emission from different depth, then summed the rotated Stokes Q and U parameters respectively to determine the resulting PA and PF.
This calculation was repeated for different wavelengths, and the results are plotted in Figure~\ref{fig:papf}.
We found that PA varies linearly with $\lambda^2$ at short wavelengths, but starts to deviate significantly at frequencies $\lesssim2$\,GHz. 
For the PF, it drops continuously as wavelength increases, and reaches half of its maximum value at $\sim$2\,GHz.
Our result suggests that the internal Faraday effects on the PA and PF are minimal at our observation frequency,
but could be significant in the L band.
For instance, we cannot simply derive the intrinsic PA using a single RM value, but need to combine it with higher-frequency (e.g., C-band) data
to solve for the Faraday dispersion function using RM synthesis \citep{rm_syn}.

\section{Conclusion} \label{sec:conclusion}
We present a polarimetric study of the PWN \pwn\ using archival C-band VLA data, and construct a simple magnetic field model to explain the observations.
Our main discovery and conclusions are summarized below:

This is the first time that the RM map of \pwn\ has been obtained.
We found symmetric structure about the pulsar spin axis, suggesting significant Faraday rotation within the PWN.
We suggest that this is due to electrons from the SN ejecta instead of pulsar wind.
In addition, the spatial variation of the RM is more likely to be caused by nonuniformly distributed ejecta electrons, rather than by the variation of the magnetic field.
Such nonuniform distribution could be the result of asymmetric SN explosion or a latitude-dependent mixing process between the pulsar wind and the ejecta.


Our polarization results reveal a global radial magnetic field for \pwn.
We constructed a simple model with a radial field plus small-scale turbulence and showed that it can generally reproduce the observed surface brightness and PF maps.
The magnetic turbulence is important in explaining the radially increasing PF.

The upcoming X-ray polarimeters, including the IXPE \citep{ixpe}, the XIPE \citep{xipe} and the eXTP \citep{extp}, can provide complementary information about the magnetic field configuration of \pwn.

\begin{acknowledgments}

C.-Y. Ng is supported by a GRF grant of the Hong Kong Government under HKU 17301618. N.~Bucciantini acknowledges financial support from the Accordo Attuativo ASI-INAF n. 2017-14-H.0 `On the escape of cosmic rays and their impact on the background plasma.' We also thank the anonymous  referee  for  the helpful  comments  and
suggestions that improved the paper.
\end{acknowledgments}



\appendix

\section{RM Structure Function} \label{sec:appen}

The RM structure function is defined as 
$D_{\rm RM}(\delta \theta) = \langle [{\rm RM}(\theta)- {\rm RM}(\theta+\delta\theta)]^2 \rangle_\theta$, 
where $\theta$ is the observing position, $\delta \theta$ is 
the angular separation, and $\langle\rangle_\theta$ means the average over all possible positions of $\theta$. We aim to find a theoretical structure function that is applicable to \pwn. \citet{minter+96} derived the RM structure function based on the Kolmogorov turbulence model:
\begin{equation}
\begin{split}
    D_{\rm RM} (\delta \theta)= &\Bigg\{ 251.226 \Bigg[ \left( \frac{n_{e}}{0.1\,{\rm cm}^{-3}} \right)^2 \left( \frac{C_B^2}{10^{-13}\,{\rm m}^{-2/3}\,\mu{\rm G}^2} \right) \\ 
    &+ \left( \frac{B_{\|}}{\mu{\rm G}} \right)^2 \left( \frac{C_n^2}{10^{-3}\,{\rm m}^{-20/3}} \right) \Bigg] \\
    &+ 23.043 \left( \frac{C_n^2}{10^{-3}\,{\rm m}^{-20/3}} \right) \\
    &\times \left( \frac{C_B^2}{10^{-13}\,{\rm m}^{-2/3}\,\mu{\rm G}^2} \right) \left( \frac{l_0}{\rm pc} \right)^{2/3} \Bigg\} \\
    &\times \left( \frac{L}{\rm kpc} \right)^{8/3} \left( \frac{\delta\theta}{\rm deg} \right)^{5/3},
\end{split}
\end{equation}
where $n_e$ is the mean electron density, $B_\|$ is the mean magnetic field strength of the line-of-sight component, $l_0$ is the outer scale of the Kolmogorov turbulence, $L$ is the distance to the target, and $C_n^2$ and $C_B^2$ indicate the levels of the fluctuations.
In particular, $C_B^2 = 5.2B^2_{\rm ran}(l_0)^{-2/3} \times 10^{-13}\,\mu{\rm G}^2\,{\rm m}^{-2/3}$, where $B_{\rm ran}$ is the strength of the random $B$-field.
Based on the observed dispersion measure DM $\approx n_e L=170\,$pc\,cm$^{-3}$ \citep{j1833_06},
RM $\approx 0.81 n_e B_\| L=110\,$rad\,m$^{-2}$ and distance $L=4.7\,$kpc of \pwn,
we estimate that $n_e = 0.036\,$cm$^{-3}$ and $B_\| = 0.8\,\mu$G.
For $l_0$ and $C^2_n$, we adopt their values from \citet{minter+96} and note that these are insensitive to our final result.
Finally, $B_{\rm ran}$ may take a wide range, from 1\,$\mu$G \citep{minter+96} to 9\,$\mu$G \citep{haverkorn+08}, so we leave it as a free parameter.
All these values are listed in Table~\ref{tab:input}. 
We stress that these are likely overestimates, since our goal is to derive an upper limit of the theoretical RM fluctuation.
\begin{table}[b!]
    \centering
    \caption{Input parameters of $D_{\rm RM}$} \label{tab:input}
    \begin{tabular}{p{4cm} c}
         \hline
         $n_e$ (cm$^{-3}$) \dotfill & 0.036 \\
         $B_\|$ ($\mu$G)\dotfill & 0.8 \\
         $L$ (kpc) \dotfill & 4.7 \\
         $l_0$ (pc) \dotfill & 3.6 \\
         $C_n^2$ ($10^{-3}\,{\rm m}^{-20/3}$) \dotfill & 1 \\
         $C_B^2$ ($10^{-13}\,{\rm m}^{-2/3}\,\mu{\rm G}^2$) \dotfill & $2.2\sim180$ \\
         $B_{\rm ran}$ ($\mu$G) \dotfill & $1\sim9$ \\
         \hline
    \end{tabular}
\end{table}
Putting these together, the final structure function is then:
\begin{equation}
    D_{\rm RM}(\delta \theta) \approx \Bigg[ 13 \left( \frac{B_{\rm ran}}{\mu{\rm G}} \right)^2 +11 \Bigg] 
    \left( \frac{\delta\theta}{1'} \right)^{5/3} \,{\rm rad}^2\,{\rm m}^{-4}.
\end{equation}



\bibliography{pwn} 
\bibliographystyle{aasjournal}

\end{document}